\documentclass[prd,tightten,aps,nofootinbib,showpacs,preprintnumbers]{revtex4}
\begin{document}
\title{Born-Infeld Electrodynamics and Euler-Heisenberg-like Model: outstanding
examples of the lack of commutativity among quantized truncated actions and
truncated quantized actions}
\author{Antonio Accioly}
\email{accioly@cbpf.br} \affiliation{Laborat\'orio de F\'{\i}sica Experimental (LAFEX),
Centro Brasileiro de Pesquisas F\'{\i}sicas (CBPF), Rua Dr. Xavier Sigaud 150, Urca,
22290-180, Rio de Janeiro, RJ, Brazil\\
Instituto de F\'{\i}sica Te\'{o}rica (IFT), S\~ao Paulo State
University (UNESP), Rua Dr. Bento Teobaldo Ferraz, 271, Bl. II-Barra
Funda, 01140-070 S\~ao Paulo, SP, Brazil}
\author{Patricio Gaete}
\email{patricio.gaete@usm.cl} \affiliation{Departmento de
F\'{\i}sica and Centro Cient\'{\i}fico-Tecnol\'{o}gico de
Valpara\'{\i}so,Universidad T\'ecnica Federico Santa Mar\'{\i}a,
Valpara\'{\i}so, Chile}
\author{Jos\'{e} A. Hela\"{y}el-Neto}
\email{helayel@cbpf.br} \affiliation{Centro Brasileiro de Pesquisas
F\'{\i}sicas, Rua Xavier Sigaud, 150, Urca, 22290-180, Rio de
Janeiro, Brazil}
\date{\today}

\begin{abstract}
We calculate the lowest-order corrections to the static potential
for both the generalized Born-Infeld Electrodynamics and an
Euler-Heisenberg-like model, in the presence of a constant external
magnetic field. Our analysis is carried out within the framework of the
gauge-invariant but path-dependent variables formalism. The
calculation reveals a long-range correction ($
{\raise0.7ex\hbox{$1$} \!\mathord{\left/ {\vphantom {1 {r^5
}}}\right.\kern-\nulldelimiterspace} \!\lower0.7ex\hbox{${r^5
}$}}$-type) to the Coulomb potential for the generalized Born-Infeld
Electrodynamics. Interestingly enough, in the Euler-Heisenberg-like
model, the static potential remains Coulombian. Therefore, contrary
to popular belief, the quantized truncated action and the truncated
quantized action do not commute at all.
\end{abstract}
\pacs{11.10.Ef, 11.15.-q}
\maketitle

\section{Introduction}

The photon-photon scattering of Quantum Electrodynamics  (QED) and
its physical consequences such as vacuum birefringence and vacuum
dichroism have been the focus of great interest for many authors
\cite{Adler,Constantini,Biswas,Tommasini,Ferrando,Michinel,Kruglov}.
As well-known, this subject has had a revival after recent
results of the PVLAS collaboration \cite{Zavattini,Bregnant}.
Although these searches have ultimately yielded upper limits on the
photon-photon cross section, the issue remains as relevant as ever.
Mention should be made, at this point, to alternative scenarios
such as Born-Infeld theory \cite{Born}, millicharged particles
\cite{Gies} or axion-like particles \cite{Masso,GaeteGuen1,GaeteSpa}
in order to account for the results reported by the PVLAS collaboration.
Also, the study of space-time noncommutativity on light propagation
in a background electromagnetic field has been discussed along these
lines \cite{Chaichian,Pinzul,Mariz}.

In this perspective, it should be mentioned that recently
considerable attention has been paid to the study of non-linear
electrodynamics (Born-Infeld theory) due to its natural emergence
from D-brane physics \cite{Tseytlin,Gibbons}. In addition to the
string interest, Born-Infeld theory has also been investigated from
quite different viewpoints. For example, in connection with
duality symmetry \cite{Parra,Rivelles,Banerjee}; also, in magnetic
monopoles studies \cite{Kim} and Lorentz symmetry breaking \cite{Botta},
by analyzing the equivalence/nonequivalence between the $\theta$-expanded
version of the noncommutative $U(1)$ gauge theory and the Born-Infeld action up
to order $F^3$ \cite{GaeteSchmidt}, and in showing the equivalence
between a $(2+1)$-dimensional topologically massive Born-Infeld
theory and $QED_3$ with a Thirring interaction term among fermions,
in the short-distance regime \cite{Gaete2004}. More recently, it has
been adopted to analize the electric and magnetostatic fields
generated by a pointlike electric charge at rest in an inertial
frame, showing a rich internal structure for the charge
\cite{Helayel}.

We further notice that very recently a generalized Born-Infeld
electrodynamics with two parameters has been investigated
\cite{Kruglov09}, which contains to leading order an
Euler-Heisenberg-like model as a particular case. It was shown
that, for this new model, the birefringence phenomenon is present.
At this point it is worth noticing that the expansion in
small parameters for generalized Born-Infeld electrodynamics is not
always smooth, that is, results for generalized Born-Infeld electrodynamics
do not always reduce to the Euler-Heisenberg-like model. At first sight
this seems to be an unexpected result since both theories are related
through an expansion in small parameters. The difference actually appears
in the quantum theory, since the quantization of a truncated action and
the truncation of the corresponding quantized action are not, of course,
commutative operations; they yield, as we should expect, different results.
We explicitly show that this is the case in the present work.

Given the outgoing experiments related to this type of physics
\cite{Mourou,Hercules}, it is desirable to have some additional
understanding of the physical consequences presented by generalized
Born-Infeld Electrodynamics. Of special interest will be to study
the connection or equivalence between generalized Born-Infeld
Electrodynamic and an Euler-Heisenberg-like model. The main goal of
this work shall be the investigation of the consequences of including two
parameters on the physical observables of Born-Infeld Electrodynamics.
In particular, the static potential between two charges, using the
gauge -invariant but path-dependent variables formalism, which is
an alternative to the Wilson loop approach. More specifically, we shall
calculate the lowest-order correction to the Coulomb energy of a
fermion-antifermion system, for both an Euler-Heisenberg-like model
and generalized Born-Infeld Electrodynamics, in an external magnetic
field. As a result, we shall show that the corrections to the static
potential obtained from these theories are quite different. This
means that the two theories are not equivalent. Generalized
Born-Infeld Electrodynamics has a rich structure which is reflected
in a long-range correction to the Coulomb potential. This correction
is identical to that encountered in Born-Infeld theory. However, in
the Euler-Heisenberg theory, the nature of the potential remains
unchanged. Accordingly, the gauge-invariant but path-dependent
variables formalism offers an alternative method in which some
features of gauge theories become more transparent.

\section{Euler-Heisenberg-like model}

Let us start off our considerations by computing the
interaction energy between static point-like sources for an
Euler-Heisenberg-like model, in an external magnetic
field. We stress that we shall restrict our computations to
the lowest order. Notice, however that, if we go over to the next order,
an $L^{-4}$-correction to the Coulomb potential shows up.
To this end, we shall compute the expectation value of the
energy operator $H$ in the physical state $|\Phi\rangle$, which
we will denote by ${\langle H\rangle}_\Phi$. The initial point of our
analysis is the following four-dimensional space-time Lagrangian
\cite{Kruglov}:
\begin{equation}
{\cal L} =  - \frac{1}{4}F_{\mu \nu } F^{\mu \nu }  + a \left(
{F_{\mu \nu } F^{\mu \nu } } \right)^2  + b \left( {F_{\mu \nu }
{\cal F}^{\mu \nu } } \right)^2, \label{EH05}
\end{equation}
where $\frac{1}{4}F_{\mu \nu } F^{\mu \nu }  =  - \frac{1}{2} \left(
{{\bf E}^2  - {\bf B}^2 } \right)$, $F_{\mu \nu } {\cal F}^{\mu \nu
} = 4\left( {{\bf E} \cdot {\bf B}} \right)$; $\cal F$ is the dual of
the field-strength $F$. While $a$ and $b$
are free parameters.  We mention, in passing, that Eq.(\ref{EH05}), in
the case of  QED (Euler-Heisenberg Lagrangian density), has the
parameters $a  = \frac{{8\alpha ^2 \hbar ^3
}}{{45m_e^4 c^5 }}$ and $b  = \frac{{14\alpha ^2 \hbar ^3
}}{{45m_e^4 c^5 }}$. Here, $\alpha$ is the fine structure constant,
and $m_e$ is the electron mass.

Next, after splitting $F_{\mu \nu }$ in the sum of a classical
background, $\left\langle {F_{\mu \nu } } \right\rangle$, and a small
fluctuation, $f_{\mu \nu }$, the corresponding Lagrangian density up
to quadratic terms in the fluctuations, is given by
\begin{equation}
{\cal L} =  - \frac{1}{4}f_{\mu \nu } \left( {1 - 2a v^{\rho
\lambda } v_{\rho \lambda }} \right)f^{\mu \nu }  + 4a f^{\rho
\lambda } \left\langle {F_{\rho \lambda } } \right\rangle
\left\langle {F_{\mu \nu } } \right\rangle f^{\mu \nu }  + b
f_{\mu \nu } v^{\mu \nu } v_{\rho \lambda } f^{\rho \lambda },
\label{EH10}
\end{equation}
where $f_{\mu \nu }  = \partial _\mu  a_\nu   - \partial _\nu  a_\mu$;
$a_\mu$ stands for the fluctuation in the potential. For the sake of
simplicity we have set
$\varepsilon^{\mu \nu \alpha \beta } \left\langle {F_{\alpha \beta }
} \right\rangle \equiv v^{\mu\nu}$ and $\varepsilon ^{\rho \lambda
\gamma \delta } \left\langle {F_{\gamma \delta } } \right\rangle
\equiv v^{\rho \lambda } $.

As already stated, our main objective will be the calculation of the
interaction energy in the $v^{0i}  \ne 0$ and $v^{ij} = 0$ case
(referred to as the magnetic one in what follows). In such a case,
the Lagrangian density (\ref{EH10}) reads
\begin{equation}
{\cal L} =  - \frac{1}{4}f_{\mu \nu } \left( {1 + 4a v^2} \right)
f^{\mu \nu }  + a \varepsilon _{\rho \lambda i0} v^{i0} f^{\rho
\lambda } \varepsilon _{\mu \nu j0} v^{j0} f^{\mu \nu }  + 4b
f_{i0} v^{i0} f_{j0} v^{j0} . \label{EH15}
\end{equation}
Having characterized the theory under study, we can now compute the
interaction energy. In this vein, we now carry out a Hamiltonian
analysis of this theory. The canonical momenta are found to be  $
\Pi ^\mu   = \left( {1 + 4a v^2} \right)f^{\mu 0} + 8b v^{0\mu }
v^{0k} f_{0k} $, which produces the usual primary constraint $\Pi ^0
= 0$ while the other momenta are $\Pi _i  = D_{ij} E_j$. Here $E_i \equiv
f_{i0}$ and $ D_{ij}  = \Omega \delta _{ij} + 8b v_{i0} v_{j0}$,
with $\Omega  = \left( {1 + 4a v^2} \right)$. Since ${\bf D}$ is
nonsingular, there exists its inverse, ${\bf D}^{ - 1} $. Thus, the
corresponding electric field takes the form
\begin{equation}
E_i  = \frac{1}{{\Omega \det D}}\left\{ {\delta _{ij} \det D -
\frac{{8b }}{\Omega }v_{i0} v_{j0} } \right\}\Pi _j , \label{EH20}
\end{equation}
where $\det D = 1 + \frac{{8b v^2 }}{\Omega}$, and $v^2  = 4{\cal
B}^2$. Here, ${\cal B}$ stands for the classical background magnetic
field around which the $a^{\mu}$-field fluctuates.

The canonical Hamiltonian is now obtained in the usual way via a
Legendre transform. It then reads
\begin{equation}
H_C  = \int {d^3 x} \left\{ { - a_0 \partial _i \Pi ^i  - \frac{1}
{{2\left[ {1 + 4\left( {a + 2b} \right){\bf v}^2 } \right]}}\Pi _i \Pi ^i  +
\frac{{\left( {1 + 4a{\bf v}^2 } \right)}}{2}{\bf B}^2  - \frac{a}{4}
\left( {{\bf v} \cdot {\bf B}} \right)^2 } \right\}, \label{EH25}
\end{equation}
where, ${\bf E}$ and ${\bf B}$ represent, respectively, electric
and magnetic fields.

From Eq. (\ref{EH25}) it is straightforward to see that the preservation
in time of the primary constraint leads to the secondary constraint,
$\Gamma_1 \left(x \right) \equiv \partial_i \Pi ^i=0$. But, the time
stability of the secondary constraint does not induce further constraints.
Therefore, the extended Hamiltonian that generates translations in time
is $H = H_C + \int {d^3 }x\left( {u_0 \left( x \right)\Pi _0 \left( x \right) +
u_1 \left( x\right)\Gamma _1 \left( x \right)} \right)$. Here, $u_0
\left( x\right)$ and $u_1 \left( x \right)$ are arbitrary Lagrange
multipliers. It should be noted that $\dot{a}_0 \left( x \right)=
\left[ {a_0 \left( x \right),H} \right] = u_0 \left( x \right)$,
which is an arbitrary function. Since $ \Pi^0 = 0$ always, neither $
a^0 $ nor $ \Pi^0 $ are of interest in describing the system and may
be discarded from the theory. Therefore, the Hamiltonian of Eq. (\ref{EH25})
can be cast under the form:
\begin{equation}
H = \int {d^3 x} \left\{ { w(x) \partial _i \Pi ^i  - \frac{1}
{{2\left[ {1 + 4\left( {a + 2b} \right){\bf v}^2 } \right]}}\Pi _i \Pi ^i  +
\frac{{\left( {1 + 4a{\bf v}^2 } \right)}}{2}{\bf B}^2  - \frac{a}{4}
\left( {{\bf v} \cdot {\bf B}} \right)^2 } \right\}, \label{EH30}
\end{equation}
where we have absorbed $(-a_0)$ appearing in (\ref{EH25}) in
$w(x) = u_1 (x) - a_0 (x)$.

In accordance with the Dirac method, we must fix the gauge. A
particularly convenient gauge-fixing condition is
\begin{equation}
\Gamma _2 \left( x \right) \equiv \int\limits_{C_{\xi x} } {dz^\nu }
a_\nu \left( z \right) \equiv \int\limits_0^1 {d\lambda x^i } a_i
\left( {\lambda x} \right) = 0, \label{EH35}
\end{equation}
where  $\lambda$ $(0\leq \lambda\leq1)$ is the parameter describing
the spacelike straight path $ x^i = \xi ^i  + \lambda \left( {x -
\xi } \right)^i $, and $ \xi $ is a fixed point (reference point).
There is no essential loss of generality if we restrict our
considerations to $ \xi ^i=0 $. The choice (\ref{EH35}) leads to the
Poincar\'e gauge \cite{GaeteZ,GaeteSPRD}. By means of this procedure,
we arrive at the only nonvanishing equal-time Dirac bracket for the
canonical variables
\begin{equation}
\left\{ {a_i \left( x \right),\Pi ^j \left( y \right)} \right\}^ *
=\delta{ _i^j} \delta ^{\left( 3 \right)} \left( {x - y} \right) -
\partial _i^x \int\limits_0^1 {d\lambda x^j } \delta ^{\left( 3
\right)} \left( {\lambda x - y} \right). \label{EH40}
\end{equation}

After achieving the quantization, we may now proceed to determine the
interaction energy for the model under consideration. To do this, we
will work out the expectation value of the energy operator $H$ in
the physical state $|\Phi\rangle$. We also recall that the physical
states $|\Phi\rangle$ are gauge invariant \cite{Dirac}. In that case
we consider the stringy gauge-invariant state
\begin{equation}
\left| \Phi  \right\rangle  \equiv \left| {\overline \Psi  \left(
\bf y \right)\Psi \left( {\bf 0} \right)} \right\rangle = \overline
\psi \left( \bf y \right)\exp \left( {iq\int\limits_{{\bf 0}}^{\bf
y} {dz^i } a_i \left( z \right)} \right)\psi \left({\bf 0}
\right)\left| 0 \right\rangle, \label{EH45}
\end{equation}
where $\left| 0 \right\rangle$ is the physical vacuum state and the
integral is to be over the linear spacelike path starting at $\bf 0$
and ending at $\bf y$, on a fixed time slice. Note that the strings
between fermions have been introduced to have a gauge-invariant
state $|\Phi\rangle$, in other terms, this means that the fermions
are now dressed by a cloud of gauge fields.

Next, taking into account the preceding Hamiltonian analysis, we then
easily verify that
\begin{equation}
\left\langle H \right\rangle _\Phi   = \left\langle H \right\rangle
_0 + \left\langle H \right\rangle _\Phi ^{\left( 1 \right)},
\label{EH50}
\end{equation}
where $\left\langle H \right\rangle _0  = \left\langle 0
\right|H\left| 0 \right\rangle$. The $\left\langle H
\right\rangle _\Phi ^{\left( 1 \right)}$-term is given by
\begin{equation}
\left\langle H \right\rangle _\Phi ^{(1)}  = \left\langle \Phi
\right|\int {d^3 x} \left\{ { - \frac{1}{{2\left[ {1 + 4\left( {a + 2b}
\right){\bf v}^2 } \right]}}\Pi _i \Pi ^i } \right\}\left| \Phi
\right\rangle, \label{EH55}
\end{equation}

Following an earlier procedure \cite{Gaete}, we see that the potential
for two opposite charges located at ${\bf 0}$ and $\bf y$ takes the form
\begin{equation}
V =  - \frac{{q^2 }}{{4\pi \left[ {1 + 16\left( {a + 2b} \right){\cal B}^2 }
\right]}}\frac{1}{L} =  - \frac{{q_{eff}^2 }}{{4\pi }}\frac{1}{L}, \label{EH60}
\end{equation}
where $|{\bf y}|\equiv L$.
Accordingly, the introduction of the external magnetic field induces a
charge redefinition. In other words, $q_{eff}$ stands for the charge
redefined upon the incorporation of the factor
$\left[ {1 + 16\left( {a + 2b} \right){\cal B}^2 } \right]^{-1}$.

It is worth noting here that there is an alternative but equivalent
way of obtaining the results (\ref{EH60}). To show this, we consider
\cite{GaeteZ}
\begin{equation}
V \equiv q\left( {{\cal A}_0 \left( {\bf 0} \right) - {\cal A}_0
\left( {\bf y} \right)} \right), \label{EH65}
\end{equation}
where the physical scalar potential is given by
\begin{equation}
{\cal A}_0 \left( {x^0 ,{\bf x}} \right) = \int_0^1 {d\lambda } x^i
E_i \left( {\lambda {\bf x}} \right), \label{EH70}
\end{equation}
with $i=1,2,3$. This equation follows from the
vector gauge-invariant field expression \cite{GaeteSPRD}
\begin{equation}
{\cal A}_\mu  \left( x \right) \equiv A_\mu  \left( x \right) +
\partial _\mu  \left( { - \int_\xi ^x {dz^\mu  } A_\mu  \left( z
\right)} \right), \label{EH75}
\end{equation}
where, as in Eq.(\ref{EH45}), the line integral is along a spacelike
path from the point $\xi$ to $x$, on a fixed slice time. It should
be noted that the gauge-invariant variables (\ref{EH75}) commute
with the sole first constraint (Gauss law), confirming in this way
that these fields are physical variables \cite{Dirac}.

Having made these observations, we see that Gauss' law for the
present theory (obtained from the Hamiltonian formulation above)
leads to $\partial _i \Pi ^i  = J^0$, where we have included the
external current $J^0$ to represent the presence of two
opposite charges. For $J^0 \left( {t,{\bf x}} \right) = q\delta
^{\left( 3 \right)} \left( {\bf x} \right)$, the electric field
then becomes
\begin{equation}
E_i  = \frac{q}{{ [{1 + 4(a+2b) {\bf v}^2 } ]}}\partial _i G\left(
{\bf x} \right), \label{EH80}
\end{equation}
where $G\left( {\bf x} \right) = \frac{1}{{4\pi }}\frac{1}{{|{\bf
x}|}}$ is the Green's function. Using this result, the physical scalar
potential, Eq.(\ref{EH70}), takes the form
\begin{equation}
{\cal A}_0 \left( {\bf x} \right) = \frac{q}{{[ {1 + 4(a+2b) {\bf v}^2 }]}}G
\left( {\bf x} \right), \label{EH85}
\end{equation}
after substraction of self-energy terms. This, together with
Eq.(\ref{EH65}), yields finally
\begin{equation}
V =  - \frac{{q_{eff}^2 }}{{4\pi }}\frac{1}{L}, \label{EH90}
\end{equation}
for a pair of point-like opposite charges q located at ${\bf 0}$ and
${\bf L}$, with $\left| {\bf L} \right| \equiv L$. It must be clear
from this discussion that a correct identification of physical degrees of
freedom is a key feature for understanding the physics hidden in
gauge theories. According to this viewpoint, once that
identification is made, the computation of the potential is carried out
by means of Gauss law \cite{Haagensen} and the effect of the external
uniform magnetic field amounts to a (finite) redefinition of the
electric field.

Before concluding this Section, we should comment on our result. Had we
considered the quartic terms as perturbations of the Maxwell Lagrangian,
a correction of the $\frac{1}{r^5}$-type would appear in the potential.
However, since we truncate our expression in the quantum fluctuations
at the $2$-nd order in the quantum fields, the Coulombian form of the
potential is not affected; the correction is only a neat redefinition
of the electric charge. We shall come back to this point at the end of
the next Section.

\section{Generalized Born-Infeld theory}

We now pass to the calculation of the interaction energy between
static pointlike sources for generalized Born-Infeld
Electrodynamics in an external background magnetic field. In other
words, we wish to explore the effects of including  two parameters
in the Born-Infeld theory on the nature of the potential. The
corresponding theory is governed by the Lagrangian density
\cite{Kruglov09}:
\begin{equation}
{\cal L} = \beta ^2 \left[ {1 - \sqrt {1 + \frac{1}{{2\beta ^2
}}F_{\mu \nu }^2 - \frac{1}{{16\beta ^2 \gamma ^2 }}\left( {F_{\mu
\nu } {\cal F}^{\mu \nu } } \right)^2 } } \right]. \label{BI5}
\end{equation}

Now, in order to handle the square root in (\ref{BI5}), we incorporate
an auxiliary field $\xi$ such that its equation of motion gives back
the original theory \cite{GaeteSchmidt}. This allows us to write the Lagrangian
density as
\begin{equation}
{\cal L} = \beta ^2 \left\{ {1 - \frac{\xi}{2}\left( {1 + \frac{1}
{{2\beta ^2 }}F_{\mu \nu } F^{\mu \nu }  - \frac{1}{{16 \beta ^2 \gamma ^2 }}
\left( {F_{\mu \nu } {\cal F}^{\mu \nu } } \right)^2 } \right) -
\frac{1}{{2\xi}}} \right\}.  \label{BI10}
\end{equation}
Notice that we are not truncating the action (\ref{BI5}). Actually,
if we expand the Lagrangian density (\ref{BI5}) up to quadratic
order in the small fluctuations, we reproduce the results of Section
II, for the Euler-Heisenberg-type model given by (\ref{EH05}), which
may be generated from the expansion of (\ref{BI5}) by keeping the
terms $F_{\mu\nu} ^2$ and $(F_{\mu\nu}{\cal F}_{\mu\nu})^2$. The
procedure of introducing the auxiliary field, $\xi$, is just because
we wish to consider the full action (\ref{BI5}). Since the
$\xi$-field is an auxiliary one, it can be readily eliminated by
means of its (algebraic) field equation. In so doing, we get
\begin{equation}
\xi  = \frac{1}{{\sqrt {1 + \frac{1}{{2\beta ^2 }}F_{\mu \nu }^2  -
\frac{1}{{16\beta ^2 \gamma ^2 }}\left( {F_{\mu \nu } {\cal F}^{\mu \nu } }
\right)^2 } }}, \label{B10b}
\end{equation}
and using it we recover eq.(\ref{BI5}).

Now, proceeding as before, after splitting $F_{\mu \nu }$ in the sum
of a classical background $\left\langle {F_{\mu \nu } }
\right\rangle$ and a small fluctuation  $f_{\mu \nu }$, we get the
corresponding Lagrangian density up to quadratic terms in the
fluctuations, namely,
\begin{equation}
{\cal L} = \beta ^2  - \frac{{\beta ^2 }}{2}\xi -
\frac{\xi}{4}f_{\mu \nu } f^{\mu \nu }  + \frac{\xi}{{32\gamma ^2
}}v^{\mu \nu } f_{\mu \nu } v^{\lambda \rho } f_{\lambda \rho }  -
\frac{{\beta ^2 }}{{2\xi}},  \label{BI15}
\end{equation}
where $\varepsilon^{\mu \nu \alpha \beta } \left\langle {F_{\alpha
\beta } } \right\rangle \equiv v^{\mu\nu}$ and $\varepsilon ^{\rho
\lambda \gamma \delta } \left\langle {F_{\gamma \delta } }
\right\rangle \equiv v^{\rho \lambda } $. The Lagrangian density
expressed by (\ref{BI15}) describes the effective dynamics of the
quantum $a_\mu$-field.

As stated previously, we are interested in the situation with $v^{0i} \ne
0$ and $v^{ij} = 0$ (referred to as the magnetic one). This
leads to the Lagrangian density
\begin{equation}
{\cal L} = \beta ^2  - \frac{{\beta ^2 }}{2}\xi -
\frac{\xi}{4}f_{\mu \nu } f^{\mu \nu }  + \frac{\xi}{{8\gamma ^2
}}v^{i0} f_{i0} v^{j0} f_{j0}  - \frac{{\beta ^2 }}{{2\xi}}.
\label{BI20}
\end{equation}

The presence of $\xi$ in (\ref{BI20}) ensures that this Lagrangian
is richer that the one given in (\ref{EH15}), for $\xi$  accounts
for higher-order terms in the fluctuations. Notice that we do not
integrate over the $\xi$-field in (\ref{BI20}) (we could, for it is
an auxiliary field); we keep it because we wish to explicitly see
how it contributes to the constraints structure with respect to the
Euler-Heisenberg-type model.

It is once again straightforward to apply the gauge-invariant
formalism discussed in the preceding section. For this purpose, we
shall first carry out its Hamiltonian analysis. The canonical
momenta read $\Pi ^\mu   = \xi f^{\mu 0}  + \frac{\xi }{{8\gamma ^2
}}v^{0\mu } v^{0k} f_{0k} $. The explicit presence of $\xi$ in this
expression for $\Pi ^\mu$ ensures that the model we consider now has
a different structure of constraints, if compared with the analysis
reported in Section II, where we simply had $\Pi ^\mu   = \left( {1
+ 4a v^2} \right)f^{\mu 0} + 8b v^{0\mu } v^{0k} f_{0k}$. This means
that the Gauss laws in both cases are different, so that we should
expect a different behavior of the static potential we shall be
attaining in what follows.

As we can see, there are two primary constraints $\Pi ^0=0$, and
$\omega \equiv \frac{{\partial L}}{{\partial \dot \xi }} = 0$.  The
canonical Hamiltonian corresponding to (\ref{BI20}) is
\begin{equation}
H_C  = \int {d^3 x} \left\{ {\Pi _i \partial ^i a^0  +
\frac{1}{{2\xi }}\left( {{\bf \Pi} ^2  + \beta ^2 } \right) +
\frac{\xi }{2}\left( {{\bf B}^2  + \beta ^2 } \right) - \beta ^2  -
\frac{{\left( {{\bf v} \cdot {\bf \Pi} } \right)^2 }}{{8\xi \gamma
^2 \left( {\det D} \right)}}} \right\}. \label{BI25}
\end{equation}
Requiring that the primary constraint $\Pi^0$ be preserved in the
course of time, one obtains  the secondary constraint $\Gamma _1
\left( x \right) \equiv \partial _i \Pi ^i= 0$. Similarly, the
consistency condition  for the constraint $\omega$ yields no further
constraints and just determines the $\xi$-field,
\begin{equation}
\xi  = \frac{1}{{\sqrt {{\bf B}^2  + \beta ^2 } }}\sqrt {\left(
{{\bf \Pi} ^2 + \beta ^2 } \right) - \frac{{\left( {{\bf v} \cdot
{\bf \Pi} } \right)^2 }}{{4\gamma ^2 \det D}}}, \label{BI30}
\end{equation}
which will be used to eliminate $\xi$. Once more, the corresponding
total (first-class) Hamiltonian that generates the time evolution of
the dynamical variables is $H = H_C + \int {d^2}x\left( {u_0
\left( x \right)\Pi _0 \left( x \right) + u_1 \left(x\right)\Gamma
_1 \left( x \right)} \right)$, where $u_0 \left(x\right)$ and $u_1
\left( x \right)$ are the Lagrange multiplier fields utilized to implement
the constraints. As before, neither $a_0(x)$ nor $\Pi _0(x)$ are of
interest in describing the system and may be discarded from the
theory. As a result, the total Hamiltonian becomes
\begin{equation}
H = \int {d^3 x} \left\{ {w\left( x \right)\partial _i \Pi ^i  +
\sqrt {{\bf B}^2  + \beta ^2 } \sqrt {\beta ^2  + \frac{{{\bf \Pi}
^2 }}{{\left( {1 + {{v^2 } \mathord{\left/ {\vphantom {{v^2 }
{4\gamma ^2 }}} \right. \kern-\nulldelimiterspace} {4\gamma ^2 }}}
\right)}}}  - \beta ^2 } \right\}, \label{BI35}
\end{equation}
where $w(x) = u_1 (x) - a_0 (x)$.

Following the same steps as those of the preceding section, we impose
now a supplementary condition on the vector potential such that the full
set of constraints becomes second class. Therefore, we adopt again the
same gauge-fixing condition (\ref{EH35}) used in
the last section.  Correspondingly, the fundamental Dirac
brackets are given by (\ref{EH40}).

We now have all the information required to compute the interaction
energy between point-like sources for this theory, where a fermion is
localized at $\bf 0$ and an antifermion at $\bf y$. As before,
we will calculate the expectation value of the energy
operator $H$ in the physical state $\left| \Phi  \right\rangle$.

From the foregoing discussion, we first observe that
\begin{equation}
\left\langle H \right\rangle _\Phi   = \left\langle \Phi  \right|\int {d^3 x}
\left\{ {
\sqrt {{\bf B}^2  + \beta ^2 } \sqrt {\beta ^2  + \frac{{{\bf \Pi}
^2 }}{{\left( {1 + {{v^2 } \mathord{\left/ {\vphantom {{v^2 }
{4\gamma ^2 }}} \right. \kern-\nulldelimiterspace} {4\gamma ^2 }}}
\right)}}}  - \beta ^2 } \right\} \left| \Phi \right\rangle. \label{BI40}
\end{equation}

Hence we see that the lowest-order modification in $\cal B$ concerning the interaction
energy may be written as
\begin{equation}
\left\langle H \right\rangle _\Phi   = \frac{1}{{\left( {1 + \frac{{{\cal B}^2 }}
{{\gamma ^2 }}} \right)}}\left\langle \Phi  \right|\int {d^3 x} \left[ {\frac{1}{2}{\bf \Pi} ^2
- \frac{1}{{8\beta ^2 }}(1 + \frac{{{\cal B}^2 }}{{\gamma ^2 }})^{ - 1}
{\bf \Pi} ^4 } \right]\left| \Phi  \right\rangle.\label{BI45}
\end{equation}

Taking into account the above Hamiltonian structure, we observe that
\begin{equation}
\Pi _i \left( x \right)\left| {\overline \Psi  \left( {\bf y}
\right)\Psi \left( {\bf 0} \right)} \right\rangle  = \overline \Psi
\left( {\bf y} \right)\Psi \left( {\bf 0} \right)\Pi _i \left( {\bf
x} \right)\left| {\bf 0} \right\rangle  - q\int_{\bf 0}^{\bf y}
{dz_i } \delta ^{\left( 3 \right)} \left( {{\bf z} - {\bf x}}
\right)\left| \Phi  \right\rangle. \label{BI50}
\end{equation}
Inserting this back into (\ref{BI45}), the lowest-order modification
in $\cal B$ of the interaction energy takes the form
\begin{equation}
V =  - \frac{{q^2 }}{{4\pi }}\frac{1}{{ { \left( {1 + {\raise0.7ex\hbox{${{\cal B}^2 }$}
\!\mathord{\left/ {\vphantom {{{\cal B}^2 } {\gamma ^2
}}}\right.\kern-\nulldelimiterspace} \!\lower0.7ex\hbox{${\gamma ^2
}$}}} \right)}}}\frac{1}{L} \left( {1 - \frac{{q^2 }}{{160\pi ^2
\beta ^2 \left( {1 + {\raise0.7ex\hbox{${{\cal B}^2 }$}
\!\mathord{\left/ {\vphantom {{B^2 } {\gamma ^2
}}}\right.\kern-\nulldelimiterspace} \!\lower0.7ex\hbox{${\gamma ^2
}$}}} \right)}}\frac{1}{{L^4 }}} \right), \label{BI55}
\end{equation}
where $|{\bf y}| = L$.

It is interesting to note that this is exactly the profile obtained
when the two parameters are identical ($\beta=\gamma$)
\cite{GaeteSchmidt}. Again, generalized Born-Infeld Electrodynamics
has a rich structure, reflected in a long-range correction
to the Coulomb potential, which is not present in the
Euler-Heisenberg-like model. Also, the precise strength of the
correction would, of course, depend on the external magnetic field.
Our results also show that the external magnetic field cannot be
arbitrarily large. Actually, $\cal B \gg \gamma$ would invalidate
the description realized by the Born-Infeld action, for there would
appear enough energy for the creation of $e^+e^-$-pairs. So, the
limit for the attainment of the potential and the validity of
Eq. (\ref{BI55}) should rule out those situations for which
$\cal B \gg \gamma$.

Here, an interesting matter comes out. Since we have considered the
quantization of the full theory, without expanding in powers of
$F_{\mu\nu}$ and no truncation was done, we get a potential richer
than the one of the Euler-Heisenberg-like model. In the latter, the
quartic terms in the fields do not modify the form of the potential,
for only the terms in $\Pi_i \Pi_i$ yield the Coulombian correction
to $V$. If we had expanded and truncated the Born-Infeld action by
keeping only the terms of order $F_{\mu\nu}^4$ and then carried out
the quantization, we would obtain the result of Section $2$ with the
appropriate identifications of the constants. This is not however
what we have done. Our procedure consisted in quantizing the full
Born-Infeld action, so that the truncation after quantization is not
expected to reproduce the same result that comes from truncating and
then quantizing.

More precisely, the Coulombian potential we have attained in Section
II, from the Euler-Heisenberg action, is valid in a regime we are
bound to consider weak fields, for the action (\ref{EH05}) appears
as a truncation of the full-fledged Born-Infeld action of
Eq.(\ref{BI5}) in the low-intensity field approximation. On the
other hand, if we keep the action (\ref{BI5}) in its full
non-polynomial form, and no truncation is performed, so that large
fields can be included, the auxiliary $\xi$-field takes into account
higher powers in the fields, as Eq.(\ref{BI30}) indicates, and so
our treatment applies also in the large-field limit. So, in
(\ref{BI45}), when we take the lowest-order correction in the
classical magnetic field, fluctuations have been incorporated that
the Euler-Heisenberg action does not include. Therefore, even if we
consider the lowest-order modification in ${\cal B}$ for the
interaction energy, Euler-Heisenberg and Born-Infeld differ, for the
latter accounts for fluctuations that the former suppresses. This
becomes actually manifest in the Hamiltonian (\ref{BI35}), where the
auxiliary $\xi$-field, as given by (\ref{BI30}), has been replaced
and so endows (\ref{BI35}) with effects that have been cut away from
the Hamiltonian of Eq. (\ref{EH30}). Our final statement is that the
potential worked out in the case of Born-Infeld is richer than the
Euler-Heisenberg's just because the $\xi$-field is kept with all
powers of the electric and magnetic fields. Then, when we analyze
the interaction energy in the lowest-order in the classical magnetic
field, we implicitly take into account effects of fluctuations
thrown away by the Euler-Heisenberg Lagrangian density.

\section{Final Remarks}

To conclude, we should highlight the different behaviors of the
potentials associated to each of the models. In the Euler-Heisenberg case,
we are actually dealing with an effective model stemming from $1$-loop
corrections to the photon-photon scattering taken in the low-frequency
limit. Since it is an effective model, the energies are below a cut-off
and we are actually studying the low-energy regime of a more complete
theory. In the Born-Infeld case, the action takes into account higher-order
terms in the frequency, so that the potential of Eq. (\ref{BI55})
incorporates contributions that, in the Euler-Heisenberg model, are
truncated to keep only the $F^4$-terms. This justifies why the two
potentials have so different $L$-dependences.

\section{ACKNOWLEDGMENTS}
This work was supported in part by Fondecyt (Chile) grant 1080260
(PG). (J.H-N) expresses his gratitude to CNPq. (A. A.) is very indebted
to FAPERJ and CNPq for partial financial support.

\end{document}